\documentstyle[12pt]{article}
\makeatletter
\def\@sect#1#2#3#4#5#6[#7]#8{\ifnum #2>\c@secnumdepth
    \def\@svsec{}\else
    \refstepcounter{#1}\edef\@svsec{\csname the#1\endcsname.\hskip 1em }\fi
    \@tempskipa #5\relax
    \ifdim \@tempskipa>\z@
    \begingroup #6\relax
    \@hangfrom{\hskip #3\relax\@svsec}{\interlinepenalty \@M #8\par}
    \endgroup
    \csname #1mark\endcsname{#7}\addcontentsline
    {toc}{#1}{\ifnum #2>\c@secnumdepth \else
     \protect\numberline{\csname the#1\endcsname}\fi
           #7}\else
    \def\@svsechd{#6\hskip #3\@svsec #8\csname #1mark\endcsname
          {#7}\addcontentsline
          {toc}{#1}{\ifnum #2>\c@secnumdepth \else
     \protect\numberline{\csname the#1\endcsname}\fi
           #7}}\fi
     \@xsect{#5}}
\def\label#1{\@bsphack\if@filesw {\let\thepage\relax
   \xdef\@gtempa{\write\@auxout{\string
   \newlabel{#1}{{\thesection.\@currentlabel}{\thepage}}}}}\@gtempa
   \if@nobreak \ifvmode\nobreak\fi\fi\fi\@esphack}
\def\@eqnnum{(\thesection.\theequation)}
\def\section{\setcounter{equation}{0} \@startsection {section}{1}{\z@}{-3.5ex
   plus -1ex minus -.2ex}{2.3ex plus .2ex}{\Large\bf}}
\newcount\@minsofar
\newcount\@min
\newcount\@cite@temp
\def\@citex[#1]#2{%
\if@filesw \immediate \write \@auxout {\string \citation {#2}}\fi
\@tempcntb\m@ne \let\@h@ld\relax \def\@citea{}%
\@min\m@ne%
\@cite{%
  \@for \@citeb:=#2\do {\@ifundefined {b@\@citeb}%
    {\@h@ld\@citea\@tempcntb\m@ne{\bf ?}%
    \@warning {Citation `\@citeb ' on page \thepage \space undefined}}%
{\@minsofar\z@ \@for \@scan@cites:=#2\do {%
  \@ifundefined{b@\@scan@cites}%
    {\@cite@temp\m@ne}
    {\@cite@temp\number\csname b@\@scan@cites \endcsname \relax}%
\ifnum\@cite@temp > \@min% select the next one to list
    \ifnum\@minsofar = \z@
      \@minsofar\number\@cite@temp
      \edef\@scan@copy{\@scan@cites}\else
    \ifnum\@cite@temp < \@minsofar
      \@minsofar\number\@cite@temp
      \edef\@scan@copy{\@scan@cites}\fi\fi\fi}\@tempcnta\@min
  \ifnum\@minsofar > \z@ % some more
    \advance\@tempcnta\@ne
    \@min\@minsofar
    \ifnum\@tempcnta=\@minsofar %   Number follows previous--hold on to it
      \ifx\@h@ld\relax
        \edef \@h@ld{\@citea\csname b@\@scan@copy\endcsname}%
    \else \edef\@h@ld{\ifmmode{-}\else--\fi\csname b@\@scan@copy\endcsname}%
      \fi
    \else \@h@ld\@citea\csname b@\@scan@copy\endcsname
          \let\@h@ld\relax
  \fi % no more
\fi}%
\def\@citea{,\penalty\@highpenalty\,}}\@h@ld}{#1}}
\def\appendixname{Appendix}
\def\appendix{\par
  \def\pre@section{\appendixname{}}
  \setcounter{section}{1}
  \@addtoreset{equation}{section}
  \def\thesection{\Alph{section}}
  \def\theequation{\arabic{equation}}}

\newcounter{@sc}
\newcounter{@scp}
\newcounter{@t}
\newlength{\@x}
\newlength{\@xa}
\newlength{\@xb}
\newlength{\@y}
\newlength{\@ya}
\newlength{\@yb}
\newsavebox{\@pt}
\def\bezier#1(#2,#3)(#4,#5)(#6,#7){\c@@sc#1\relax
 \c@@scp\c@@sc \advance\c@@scp\@ne
 \@xb #4\unitlength \advance\@xb -#2\unitlength \multiply\@xb \tw@
 \@xa #6\unitlength \advance\@xa -#2\unitlength
 \advance\@xa -\@xb \divide\@xa\c@@sc
 \@yb #5\unitlength \advance\@yb -#3\unitlength \multiply\@yb \tw@
 \@ya #7\unitlength \advance\@ya -#3\unitlength
 \advance\@ya -\@yb \divide\@ya\c@@sc
 \setbox\@pt\hbox{\vrule height\@halfwidth depth\@halfwidth
 width\@wholewidth}\c@@t\z@
 \put(#2,#3){\@whilenum{\c@@t<\c@@scp}\do
 {\@x\c@@t\@xa \advance\@x\@xb \divide\@x\c@@sc \multiply\@x\c@@t
 \@y\c@@t\@ya \advance\@y\@yb \divide\@y\c@@sc \multiply\@y\c@@t
 \raise \@y \hbox to \z@{\hskip \@x\unhcopy\@pt\hss}\advance\c@@t\@ne}}}

\makeatother
\begin{document}
\addtolength{\unitlength}{-0.5\unitlength}
\def\t{\theta}
\def\ov{\overline}
\def\a{\alpha}
\def\b{\beta}
\def\g{\gamma}
\def\wt{\widetilde}
\def\w{\omega}
\def\ds{\displaystyle}
\def\s{\sigma}
\phantom{aa}
\vspace*{4cm}

\centerline{\bf New solution of vertex type tetrahedron equations
\footnote{Research partially supported by National Science Foundation
Grant PHY -- 93 -- 07 -- 816 and by International Science Foundation
(INTAS), Grant RMM000.}}
\vspace{1cm}
\centerline{
V.V. Mangazeev\footnote{E-mail: mangazeev@mx.ihep.su},
S.M. Sergeev\footnote{Branch Inst. for Nucl. Phys.,E-mail:
sergeev\_ms@mx.ihep.su}
and Yu.G. Stroganov\footnote{E-mail: stroganov@mx.ihep.su}}
\vspace{1cm}
\centerline{Institute for High Energy Physics,}
\centerline{Protvino, Moscow Region, Russia}

\vspace{1cm}
\centerline{Abstract}
In this paper we formulate a new $N$-state spin integrable model
on a three-dimensional lattice with spins interacting round
each elementary cube of the lattice.
This model can be also reformulated as a vertex type model.
Weight functions of the model satisfy tetrahedron equations.
\newpage
\section{Introduction}

Recently in Ref. \cite{Korep} Korepanov has constructed a new
solution of tetrahedron equations
with spin variables lying on the edges of a 3D cubic lattice \cite{BS}.
It leads to a commuting family of transfer-matrices
and possible integrability of this vertex model on the
cubic lattice. Corresponding spin variables take $N=2$ values.
There are only 16 nonzero weights $R_{i_1i_2i_3}^{j_1j_2j_3}$ from
64 possible ones.

Later Hietarinta  in Ref. \cite{Hiet1} has proposed another vertex solution
with 16 nonzero weights. It strongly reminds a solution of Ref. \cite{Korep},
but four  weights from sixteen in these two models are different.
This model satisfy to a duality B property in a terminology of Ref.
\cite{Hiet2} and can be reformulated as an interaction-round-cube (IRC)
model with weight function $W(a|efg|bcd|h)$ depending on eight surrounding
spin variables of an elementary cube of the lattice.
Note that Baxter in his paper \cite{B} also has reformulated
Zamolodchikov model \cite{Z} with spin variables belonging to faces of
3D lattice as an IRC model. We fail to generalize the original Korepanov
solution for an arbitrary $N$.

In this paper we propose a generalization of Hietarinta solution
of tetrahedron equations for an arbitrary number $N$ of spin variables.
The obtained solution has a simple multiplicative form, depends on
four parameters, which have a simple geometric interpretation.
Namely, one can choose as parameters of our solution four angles of
an arbitrary quadrilateral with two diagonals.
Hietarinta solution appears in the case if this quadrilateral can be
inscribed in the circle and $N=2$.
Also note that in the case $N=2$ our solution can be obtained as
a particular case of Zamolodchikov model \cite{B,Z}.

The paper is organized as follows. In section 2 we recall
a formulation of IRC type models , introduce necessary definitions
and  propose an ansatz for weight functions. In section 3 we show that
the tetrahedron equations are reduced for this case to some special
identity, which is proved in appendix. Section 4 is devoted to
a parameterization of obtained $N$-state solution. And, at last,
in section 5 we discuss briefly results of this paper and its
possible generalizations.

\section{Formulation of The Model}

In this section we recall some definitions following Refs. \cite{BB,KMS}.
Consider a simple cubic lattice ${\cal L}$ and at each
site of ${\cal L}$ place a spin variable taking its values in $Z_N$,
for any integer $N\ge2$
(elements of $Z_N$ are given by $N$ distinct numbers $0,1,\ldots,N-1$
considered modulo $N$).
 Allow all possible interactions of the spins within each
elementary cube. The partition function reads
\begin{equation}
Z=\sum_{spins}\prod_{cubes}W(a|e,f,g|b,c,d|h),                  \label{f1}
\end{equation}
where $W(a|e,f,g|b,c,d|h)$ is the Boltzmann weight of the spin configuration
$a,\ldots,h$.

We need in some notations to formulate our ansatz for the Boltzmann
weights.  Denote
\begin{equation}
\w=\exp(2\pi i/N),\qquad \w^{1/2}=\exp(\pi i/N).                \label{f2}
\end{equation}

Taking $x,y,z$ to be complex parameters constrained by the
Fermat equation
\begin{equation}
x^N+y^N=z^N                                                    \label{f3}
\end{equation}
and $l$ to be an element of $Z_N$, define
\begin{equation}
w(x,y,z|l)=\prod_{s=1}^{l}{y\over z-x\w^s}.                    \label{f4}
\end{equation}
This function has a following property
\begin{equation}
w(x,y,z|l)w(z,\w^{1/2}y,\w x|-l)\Phi(l)=1,\quad l\in Z_N,   \label{f5}
\end{equation}
where
\begin{equation}
\Phi(l)=\w^{l(l+N)/2}.                                          \label{f6}
\end{equation}

The tetrahedron equations can be written as (Eqs. (2.2) of Ref. \cite{B})
\begin{eqnarray}
\sum_{d}
&W(a_4|c_2,c_1,c_3|b_1,b_3,b_2|d)W'(c_1|b_2,a_3,b_1|c_4,d,c_6|b_4)&
\nonumber\\
\times&W''(b_1|d,c_4,c_3|a_2,b_3,b_4|c_5)
W'''(d|b_2,b_4,b_3|c_5,c_2,c_6|a_1)&\nonumber\\
=\sum_{d}
&W'''(b_1|c_1,c_4,c_3|a_2,a_4,a_3|d)W''(c_1|b_2,a_3,a_4|d,c_2,c_6|a_1)&
\nonumber\\
\times&W'(a_4|c_2,d,c_3|a_2,b_3,a_1|c_5)
W(d|a_1,a_3,a_2|c_4,c_5,c_6|b_4),&                              \label{2}
\end{eqnarray}
where $W$, $W'$, $W''$ and $W'''$ are some four
 sets of Boltzmann weights.
As has been shown in Refs. \cite{BS,JM}, this relation ensures
commutativity of layer-to-layer transfer matrices
constructed from $W$ and $W'$ weights:
\begin{equation}
T(W)T(W')=T(W')T(W).                                           \label{2a}
\end{equation}

We are looking for a solution for (\ref{2}) in the following form:
\begin{eqnarray}
&W(a|efg|bcd|h) =\w^{(h-f)(a-b-e+h)}\times&\nonumber\\
&\times{\ds w(p_2,p_{12},p_1|a-b-e+h)w(p_4,p_{34},p_3|-a+c+f-h)\over
 \ds w(p_6,p_{56},p_5|-b+c-e+f)}&,     \label{ch1}
\end{eqnarray}
where functions $w$ are defined by (\ref{f4}) and coordinates $p_i$, $p_{ij}$
satisfy Fermat constraint (\ref{f3}).
We imply that the set of weight functions $W'$ depend on parameters
$p'_i$, $p'_{ij}$ and etc.

\section{Proof of the Tetrahedron Equations}

In this section
we will show that tetrahedron equations (\ref{2}) for weight functions
(\ref{ch1}) are equivalent to the identity (\ref{A8}) provided that
parameters of weights satisfy some special algebraic constraints.

Substituting formula (\ref{ch1}) in (\ref{2}) we obtain
\begin{eqnarray}\label{ch2}
&{\ds w(p'_2,p'_{12},p'_1|b-a)\over\ds w(p'''_6,p'''_{56},p'''_5|c-d)}
{\ds w(p''_4,p''_{34},p''_3|c'-b')\over
\ds w(p_6,p_{56},p_5|d'-a')}\times&\nonumber\\
&{\ds\sum_{\s\in Z_N}}
w^{-1}(p'_6,p'_{56},p'_5|-a-\s)w(p'_4,p'_{34},p'_3|-b-\s)
w(p'''_4,p'''_{34},p'''_3|c+\s)&\nonumber\\
&\times w(p'''_2,p'''_{12},p'''_1|-d-\s)w(p_4,p_{34},p_3|-b'+\s)
w(p''_2,p''_{12},p''_1|-d'+\s)&\nonumber\\
&\times w(p_2,p_{12},p_1|b'+d'-a'-\s)
w^{-1}(p''_6,p''_{56},p''_5|c'-b'-d'+\s)\w^{\s^2+\s(c-c')}&\nonumber\\
&={\ds w(p''_4,p''_{34},p''_3|c-b)\over \ds w(p_6,p_{56},p_5|d-a)}
{\ds w(p'_2,p'_{12},p'_1|b'-a')\over
\ds w(p'''_6,p'''_{56},p'''_5|c'-d')}\times& \\
&{\ds\sum_{\s\in Z_N}}
w^{-1}(p'_6,p'_{56},p'_5|-a'-\s)w(p'_4,p'_{34},p'_3|-b'-\s)
w(p'''_4,p'''_{34},p'''_3|c'+\s)&\nonumber\\
&\times w(p'''_2,p'''_{12},p'''_1|-d'-\s)w(p_4,p_{34},p_3|-b+\s)
w(p''_2,p''_{12},p''_1|-d+\s)&\nonumber\\
&\times w(p_2,p_{12},p_1|b+d-a-\s)
w^{-1}(p''_6,p''_{56},p''_5|c-b-d+\s)\w^{\s^2+\s(c'-c)},&\nonumber
\end{eqnarray}
where we made a replacement $\s\to-\s$ in the LHS, $\s\to c_1+c_5-\s$ in the
RHS and introduced new spin variables:
\begin{eqnarray}
&a=-a_3+b_2+c_4,\quad &a'=a_2-b_3-c_1+c_2-c_5,\nonumber\\
&b=-a_3+b_4+c_1,\quad &b'=a_4-b_3-c_1,\nonumber\\
&c=-a_1+b_4+c_2,\quad &c'=a_4-b_1-c_1+c_4-c_5,\\
&d=-a_1+b_2+c_5,\quad &d'=a_2-b_1-c_5.\nonumber
\end{eqnarray}

Relation (\ref{ch2}) looks very similar to identity (\ref{A8}) from
appendix. But for complete coincidence of spin structure of
(\ref{ch2}) and (\ref{A8}) we need to use (\ref{f5}).
Using (\ref{f5}) several times we can easily reduce (\ref{ch2}) to
identity (\ref{A8}) with the following identification:
\begin{eqnarray}
\vec x\sim(p'_5,\w^{1/2}p'_{56},\w p'_6),&\quad
&\vec x'''\sim(p''_5,\w^{1/2}p''_{56},\w p''_6),\nonumber\\
\vec y\sim(p'_3,\w^{1/2}p'_{34},\w p'_4),&\quad
&\vec y'''\sim(p_3,\w^{1/2}p_{34},\w p_4),\nonumber\\
\vec z\sim(p'''_4,p'''_{34},p'''_3),&\quad
&\vec z'''\sim(p_2,p_{12},p_1),\nonumber\\
\vec u\sim(p'''_1,\w^{1/2}p'''_{12},\w p'''_2),&\quad
&\vec u'''\sim(p''_1,\w^{1/2}p''_{12},\w p''_2),\label{ch5}\\
(x_1y_3,\xi,\w x_3y_1)\sim(p'_1,\w^{1/2}p'_{12},\w p'_2),&\quad
&(y_3z_1,\xi'',\w y_1z_3)\sim(p''_4,p''_{34},p''_3),\nonumber\\
(z_3u_1,\tilde\xi,z_1u_3)\sim(p'''_5,\w^{1/2}p'''_{56},\w p'''_6),&\quad
&(x_3u_1,\tilde\xi'',x_1u_3)\sim(p_6,p_{56},p_5).\nonumber
\end{eqnarray}
Since we can choose four vectors $\vec x$, $\vec y$, $\vec z$ and $\vec u$
in an arbitrary way, we come to the four-parametric solution of
tetrahedron equations (\ref{2}). In the next section we will show
that this solution can be parameterized by the angles of a degenerate
tetrahedron with all apices lying in one plane.

\section{Parameterization of solution}

Using formulas (\ref{A5}-\ref{A6}) from appendix A we can exclude
parameters entering in $\vec x$, $\vec y$, $\vec z$ and $\vec u$
from relations (\ref{ch5}). Then we come to the following
constraints among parameters $p$:
\begin{equation}
{\ds p_6\over\ds p_5}={\ds p_2p_4\over\ds p_1p_3},\quad
{\ds p'_6\over\ds p'_5}={\ds p'_2p'_4\over\ds p'_1p'_3},\quad
{\ds p''_6\over\ds p''_5}={\ds p''_2p''_4\over\ds p''_1p''_3},\quad
{\ds p'''_6\over\ds p'''_5}={\ds p'''_2p'''_4\over\ds p'''_1p'''_3},\label{p1}
\end{equation}
\begin{equation}
{\ds p'_2\over\ds p'_1}={\ds p_2p''_2\over\ds p_1p''_1},\quad
{\ds p''_4\over\ds p''_3}={\ds p'_4p'''_4\over\ds p'_3p'''_3},\quad
{\ds p_4p'''_2\over \ds p_3p'''_1}={\ds p'_4p''_2\over\ds p'_3p''_1},\label{p2}
\end{equation}
\begin{equation}
{\ds p'_2p'_{56}\over\ds p'_6p'_{12}}
{\ds p''_6p''_{12}\over\ds p''_2p''_{56}}
{\ds p'''_2p'''_{56}\over\ds p'''_6p'''_{12}}=1,\>
{\ds p_1p_{56}\over\ds p_5p_{12}}
{\ds p''_3p''_{12}\over\ds p''_2p''_{34}}
{\ds p'''_2p'''_{34}\over\ds p'''_3p'''_{12}}=1,\>
{\ds p_5p_{34}\over\ds p_4p_{56}}
{\ds p'_3p'_{56}\over\ds p'_5p'_{34}}
{\ds p''_6p''_{34}\over\ds p''_3p''_{56}}
{\ds p'''_5\over\ds p'''_6}=1.   \label{p3}
\end{equation}
Also we obtain additional constraints on parameters $p_i$ corresponding
to formulas (\ref{A6}) for $\xi'^N$ and $\tilde\xi'^N$ from appendix A.
These constraints ensure a periodicity ({\it modulo} $N$) of all $w$
functions entering in (\ref{ch1}) and can be easily obtained as a
consequence of (\ref{p1}-\ref{p3}).

Remind that the function $w(p_j,p_{ij},p_i|n)$ depends on one
complex parameter $p_i/p_j$ and parameter $p_{ij}$ defined by
a relation $p_{ij}^N=p_i^N-p_j^N$ can contain an arbitrary multiplier
$\w^{n_{ij}}$, where $n_{ij}\in Z$.

It appears that instead of variables $p_i$, $p_{ij}$ and $p_j$
it is more convenient to introduce "angle-like" variables
$a_1$, $a_2$ and $a_3$ as
\begin{eqnarray}
&p_2=\exp(-ia_1/N),\quad p_1=\exp(ia_1/N),\quad p_{12}=\{2i\sin(a_1)\}^{1/N}
\w^{n_{12}},&\nonumber\\
&p_4=\exp(-ia_2/N),\quad p_3=\exp(ia_2/N),\quad p_{34}=\{2i\sin(a_2)\}^{1/N}
\w^{n_{34}},&\label{p4}\\
&p_6=\exp(-ia_3/N),\quad p_5=\exp(ia_3/N),\quad p_{56}=\{2i\sin(a_3)\}^{1/N}
\w^{n_{56}}&\nonumber
\end{eqnarray}
and etc.

Then relations (\ref{p1}-\ref{p2}) are reduced to
linear constraints among $a_i$:
\begin{eqnarray}
&a_3=a_1+a_2,\quad a'_3=a'_1+a'_2,\quad a''_3=a''_1+a''_2,\quad
a'''_3=a'''_1+a'''_2,&\nonumber\\
&a_1-a'_1+a''_1=0,\quad a'_2-a''_2+a'''_2=0,\quad
a_2-a'_2-a''_1+a'''_1=0.&\label{p5}
\end{eqnarray}

Now let us fix a choice of phase multipliers $\w^{n_{ij}}$.
Taking into account formula (\ref{ch1})  it is easy to see
that we have the following dependence of the weight function from $n_{ij}$:
\begin{equation}
\w^{n_{12}(a-b-e+h)+n_{34}(-a+c+f-h)+n_{56}(b-c+e-f)}.      \label{p6}
\end{equation}
In fact, this multiplier corresponds to a simple gauge transformation
of the weight function and hereafter we will omit all multipliers
$\w^{n_{ij}}$.

Then relations (\ref{p3}) can be rewritten in the following form:
\begin{eqnarray}
&{\ds\sin(a'_3)\over\ds\sin(a'_1)}
{\ds\sin(a''_1)\over\ds\sin(a''_3)}
{\ds\sin(a'''_3)\over\ds\sin(a'''_1)}=1,\quad
{\ds\sin(a_3)\over\ds\sin(a_1)}
{\ds\sin(a''_1)\over\ds\sin(a''_2)}
{\ds\sin(a'''_2)\over\ds\sin(a'''_1)}=1,&\nonumber\\
&{\ds\sin(a_2)\over\ds\sin(a_3)}
{\ds\sin(a'_3)\over\ds\sin(a'_2)}
{\ds\sin(a''_2)\over\ds\sin(a''_3)}=1.&\label{p7}
\end{eqnarray}

Formulas (\ref{p5},\ref{p7}) have a nice geometrical interpretation
(see Fig. 1)
\begin{picture}(600,450)
\thicklines\put(120,30){\begin{picture}(400,400)\thicklines
\put(30,150){\line(1,0){400}}
\put(30,150){\line(3,-1){277}}\put(490,195){\line(-4,-3){230}}
\put(30,150){\line(4,5){177}}\put(430,150){\line(-1,1){222}}
\put(205.3,370){\line(1,-3){104}}\bezier{50}(80,150)(77,170)(60,187.5)
\bezier{50}(100,150)(98.5,135)(90,130)\bezier{50}(174,330)(195,319.5)(218,323)
\bezier{50}(224,311)(238,309)(258.5,319)\bezier{50}(270,68)(265,52)(275,36)
\bezier{50}(300,90)(320,92)(342,85)\bezier{50}(388,150)(384,137)(397,126.5)
\bezier{50}(405,175)(430,181)(459,172)\put(0,140){$P$}\put(313,37){$P'$}
\put(435,130){$P''$}\put(205,377){$P'''$}\put(105,133){$a_1$}
\put(80,170){$a_2$}\put(241,49){$a'_1$}\put(315,100){$a'_2$}
\put(185,301){$a'''_1$}\put(239,293){$a'''_2$}\put(361,128){$a''_1$}
\put(422,188){$a''_2$}\end{picture}}
\put(310,0){\Large\bf Fig. 1}
\end{picture}
\vspace*{0.2cm}

\noindent
Then formulas (\ref{p5}) are valid in an evident way and formulas (\ref{p7})
are the consequences of sine theorem for four triangles formed by
two sides and one of the diagonals of the quadrilateral.

\section{Discussion}

In the previous section we obtained a four-parametric solution of the
tetrahedron
equations for which each weight function depends on two spectral parameters.

Namely
\begin{equation}
W=W(a_1,a_2),\quad W'=W(a'_1,a'_2),\quad W''=W(a''_1,a''_2),\quad
W'''=W(a'''_1,a'''_2).           \label{d1}
\end{equation}
Weight functions are defined by multiplicative ansatz (\ref{ch1}) where
parameters $p_i$ and $p_{ij}$ can be calculated through angle variables
$a_1$, $a_2$ from formulas (\ref{p4}) (remind that $a_3=a_1+a_2$).
All set of angle variables $a_i$, $a'_i$, $a''_i$ and $a'''_i$ is
described by Fig.~1.

It is naturally to ask: is this solution of tetrahedron equations new or not?
To clarify this situation we have considered the only known solution
of tetrahedron equations with $N > 2$
\cite{StSq} corresponding to the model
proposed by Bazhanov and Baxter  \cite{BB} in the "plane" limit for which
all apices of tetrahedron belong to one plane.
More precisely, let us consider a parameterization of Ref. \cite{KMS}
and take the limit $\t_1$, $\t_2$, $\t_4$, $\t_5\to0$ and $\t_3$, $\t_6\to\pi$.
This limit corresponds explicitly to parameterization
(\ref{d1}) and to Fig.~1. Formulas for weight functions look as
\begin{eqnarray}
&W(a|efg|bcd|h)=(-1)^{a+c+f+h}\w^{{1\over2}(-a^2-c^2+f^2+h^2+2ag-2bf)}&
\nonumber\\
&\ds\Bigl\{\sum_{\s\in Z_N}{\ds w(x_2,\w^{1/2}x_{23},x_3|b-f+\s)\over
\ds w(x_4,x_{14},x_1|e-c-\s)}\w^{\s(a-c-f+h)}\Bigr\}_0,&     \label{d2}
\end{eqnarray}
where
\begin{eqnarray}
&x_1=\exp(ia_2/N)(\sin a_1)^{1/N},\quad
x_4=\w^{-1/2}\exp(-ia_1/N)(\sin a_2)^{1/N},&\nonumber\\
&x_2=\exp(ia_1/N)(\sin a_2)^{1/N},\quad
x_3=\w^{1/2}\exp(-ia_2/N)(\sin a_1)^{1/N},&\label{d3}\\
&x_{14}=(\sin(a_1+a_2))^{1/N},\quad x_{23}=(\sin(a_1+a_2))^{1/N},&\nonumber
\end{eqnarray}
and angles $a_i$ appear as limit values of plane tetrahedron angles.
The subscript ``$0$'' after the curly brackets implies that the expression
in the curly brackets is divided by itself with all exterior spin
variables equated to zero.
If the signs of $\s$ in $w$ functions entering in (\ref{d2}) were
equal we would use formula (A14) of appendix A of Ref. \cite{StSq}
and come to a multiplicative solution.
It can be done for the case $N=2$ where all spin variables belong $Z_2$
and we can change all signs in an appropriate way. Only in this case
we obtain from (\ref{d2}) our multiplicative solution up to some
gauge transformation.

So it seems that multiplicative solution (\ref{ch1}) should be new
for $N\ge3$. Unfortunately, we did not succeed yet in a generalization
of this solution to a five-parametric one which could be described
by the angles of the usual tetrahedron.

Note also that solution (\ref{ch1}) can be naturally rewritten in terms
spin variables lying on the edges of the lattice.

Let us define the following spin variables  \cite{Hiet2}
\begin{eqnarray}
i_1=-a+c,\quad i_2=e-f,\quad i_3=a-b,&\nonumber\\
j_1=-f+h,\quad j_2=-b+c,\quad j_3=e-h.&\label{d4}
\end{eqnarray}
Then we can rewrite our weight function (\ref{ch1}) as
\begin{equation}
R^{j_1j_2j_3}_{i_1i_2i_3}=\w^{j_1(i_3-j_3)}
{\ds w(p_2,p_{12},p_1|i_3-j_3)w(p_4,p_{34},p_3|i_1-j_1)\over
 \ds w(p_6,p_{56},p_5|j_2-i_2)}.       \label{d5}
\end{equation}
Note that spin variables $i_1$, $i_2$, $i_3$, $j_1$, $j_2$, $j_3$ should
satisfy two constraints
\begin{equation}
j_2=i_1+i_3,\quad i_2=j_1+j_3.         \label{d6}
\end{equation}
Then tetrahedron equations take the vertex form
\begin{eqnarray}
&{\ds\sum_{k_1,k_2,k_3,\atop k_4,k_5,k_6}}
R^{k_1,k_2,k_3}_{i_1,i_2,i_3}R'^{j_1k_4k_5}_{\phantom{,}k_1i_4i_5}
R''^{j_2j_4k_6}_{\phantom{,,}k_2k_4i_6}
R'''^{j_3j_5j_6}_{\phantom{,,,}k_3k_5k_6}=&\nonumber\\
&={\ds\sum_{k_1,k_2,k_3,\atop k_4,k_5,k_6}}
R'''^{k_3,k_5,k_6}_{\phantom{,,,}i_3,i_5,i_6}
R''^{k_2k_4j_6}_{\phantom{,,}i_2i_4k_6}
R'^{k_1j_4j_5}_{\phantom{,}i_1k_4k_5}
R^{j_1j_2j_3}_{k_1k_2k_3}&.\label{d7}
\end{eqnarray}

At last let us suppose that the quadrilateral shown in Fig.~1 can be
inscribed in the circle. In this case all relations constrained angles
$a_i$, $a'_i$, $a''_i$ and $a'''_i$ appear to be linear,
and we have only three independent angle variables. For $N=2$
this model coincides with that of Ref. \cite{Hiet1} (to be exact,
we are to change else $a_j\rightarrow ia_j$).

{\bf Acknowledgements.}  We are indebted to J. Hietarinta
for giving an opportunity to get acquainted with his solution
prior to publication and for the valuable discussion.
Yu.G. Stroganov is also grateful to the staff of the Physics Department
of Oklahoma State University, where
a part of this work has been done,  for their kind hospitality.

\appendix
\section*{Appendix}

In this Appendix we will prove the identity (\ref{ch2}), from which it
follows the tetrahedron equation for the multiplicative ansatz.

We begin with the following relation:
\begin{eqnarray}
&\ds\left\{\sum_{\s\in Z_N}{\ds w(\vec x|a+\s) w(\vec z|c+\s)\over
\ds w(\vec y|b+\s) w(\vec u|d+\s)}\w^{\s n}\right\}_0=
\w^{-nc}
\ds {\ds w(x_1y_3,\xi,\w x_3y_1|a-b)\over\ds w(z_3u_1,\tilde\xi,z_1u_3|d-c)}
&\nonumber\\
&\ds\times\left\{\sum_{\s\in Z_N}{\ds w(\vec x'|\s) w(\vec z'|d-c+n+\s)\over
\ds w(\vec y'|n+\s) w(\vec u'|a-b+\s)}\w^{\s(b-c)}\right\}_0&\label{A1}
\end{eqnarray}
where
\begin{eqnarray}
&\vec x' = (x_3y_2,\xi,x_2y_3)&\nonumber\\
&\vec y' = (u_3z_2/\w,\tilde\xi,z_3u_2)&\nonumber\\
&\vec z' = (u_1z_2,\tilde\xi,u_2z_1)&\nonumber\\
&\vec u' = (x_1y_2,\xi,\w x_2y_1),&\label{A2}
\end{eqnarray}
the vectors $\vec x,\vec y,\vec z,\vec u$ are arbitrary, and the auxiliary
parameters $\xi$ and $\tilde\xi$ obey the relations
\begin{eqnarray}
&\xi^N = x_3^Ny_1^N-x_1^Ny_3^N&\nonumber\\
&\tilde\xi^N = z_1^Nu_3^N - z_3^Nu_1^N,&\label{A3}
\end{eqnarray}
phases of $\xi$ and $\tilde\xi$ can be chosen in arbitrary way.
The subscript ``$0$'' after the curly brackets implies that the expression
in the curly brackets is divided by itself with all the exterior spin
variables equated to zero.

This relation is equivalent to the relation (B9) from the Appendix B
of Ref. \cite{StSq} and can be interpreted as some symmetry
transformation of the Boltzmann weights of the generalization of the
Bazhanov -- Baxter model, considered in Ref. \cite{NewTw}.

Applying this transformation triple, we obtain the following relation:
\begin{eqnarray}
&\ds\left\{\sum_{\s\in Z_N}{\ds w(\vec x|a+\s) w(\vec z|c+\s)\over
\ds w(\vec y|b+\s) w(\vec u|d+\s)}\w^{\s n}\right\}_0=&\nonumber\\
&\ds\w^{-n(b+d)+d(a-b+c-d)}{\ds w(x_3y_2z_3u_2,\xi',x_2y_3z_2u_3|-n)\over
\ds w(x_1y_2z_1u_2,\tilde\xi',\w x_2y_1z_2u_1|a-b+c-d-n)}&\nonumber\\
&\ds{\ds w(x_1y_3,\xi,\w x_3y_1|a-b)\over\ds w(z_3u_1,\tilde\xi,z_1u_3|d-c)}
{\ds w(y_3z_1,\xi'',\w y_1z_3|c-b)\over\ds w(x_3u_1,\tilde\xi'',x_1u_3|d-a)}
&\nonumber\\
&\ds\left\{\sum_{\s\in Z_N}{\ds w(\vec x'''|-c+\s) w(\vec z'''|-a+\s)\over
\ds w(\vec y'''|-d+\s) w(\vec u'''|-b+\s)}\w^{\s(-a+b-c+d+n)}\right\}_0&
\label{A4}
\end{eqnarray}
where
\begin{eqnarray}
&\vec x''' = (z_3\tilde\xi',\xi''\tilde\xi x_2,z_1\xi')&\nonumber\\
&\vec y''' = (u_3\tilde\xi'/\w,\tilde\xi''\tilde\xi y_2,u_1\xi')&\nonumber\\
&\vec z''' = (x_3\tilde\xi',\tilde\xi''\xi z_2,x_1\xi')&\nonumber\\
&\vec u''' = (y_3\tilde\xi',\xi''\xi u_2,\w y_1\xi'),&\label{A5}
\end{eqnarray}
and
\begin{eqnarray}
&\xi'^N = x_2^Ny_3^Nz_2^Nu_3^N-x_3^Ny_2^Nz_3^Nu_2^N&\nonumber\\
&\tilde\xi'^N = x_2^Ny_1^Nz_2^Nu_1^N-x_1^Ny_2^Nz_1^Nu_2^N&\nonumber\\
&\xi''^N = y_1^Nz_3^N-y_3^Nz_1^N&\nonumber\\
&\tilde\xi''^N = x_1^Nu_3^N - x_3^Nu_1^N&\label{A6}
\end{eqnarray}
Note that in the special case when $x_1y_3z_1u_3=x_3y_1z_3u_1\w$
the summands from both parts of (\ref{A4}) become the Boltzmann weights
for the Bazhanov -- Baxter model  \cite{BB}, and (\ref{A4}) becomes
the well known Star -- Star relation.

Consider now the second copy of (\ref{A4}) with the same
$\vec x,\vec y,\vec z,\vec u$, with the same number $n$ and
with some spines $a',b',c',d'$ obeying
\begin{equation}
a-b+c-d = a'-b'+c'-d'.\label{A7}
\end{equation}
Multiplying left hand side of one copy of (\ref{A4}) by the right hand side
of the other copy, we obtain the identity, from which the $w$ -- functions
containing the spin $n$ are cancel out. Summing the obtained relation by
$n$, we obtain
\begin{eqnarray}
&\ds\Biggl\{\sum_{\s\in Z_N}
{\ds w(\vec x|a+\s) w(\vec z|c+\s)\over\ds w(\vec y|b+\s)w(\vec u|d+\s)}
{\ds w(\vec x'''|b'-c'+d'-\s) w(\vec z'''|b'-a'+d'-\s)
\over\ds w(\vec y'''|b'-\s)w(\vec u'''|d'-\s)}&\nonumber\\
&\times\w^{(\s-b')(a'-b'+c'-d')}\Biggl\}_0
{\ds w(x_1y_3,\xi,\w x_3y_1|a'-b')\over\ds w(z_3u_1,\tilde\xi,z_1u_3|d'-c')}
{\ds w(y_3z_1,\xi'',\w y_1z_3|c'-b')\over\ds
w(x_3u_1,\tilde\xi'',x_1u_3|d'-a')}=& \nonumber\\
&\ds\Biggl\{\sum_{\s\in Z_N}
{\ds w(\vec x|a'+\s) w(\vec z|c'+\s)\over\ds w(\vec y|b'+\s)w(\vec u|d'+\s)}
{\ds w(\vec x'''|b-c+d-\s) w(\vec z'''|b-a+d-\s)\over\ds w(\vec
y'''|b-\s)w(\vec u'''|d-\s)} &\nonumber\\
&\times\w^{(\s-b)(a-b+c-d)}\Biggl\}_0
{\ds w(x_1y_3,\xi,\w x_3y_1|a-b)\over\ds w(z_3u_1,\tilde\xi,z_1u_3|d-c)}
{\ds w(y_3z_1,\xi'',\w y_1z_3|c-b)\over\ds
w(x_3u_1,\tilde\xi'',x_1u_3|d-a)}& \label{A8}
\end{eqnarray}


\begin{thebibliography}{**}
% 1
\bibitem{Korep} I.G. Korepanov, {\it Comm. Math. Phys.} {\bf 154} (1993) 85.
% 2
\bibitem{BS}
V.V. Bazhanov, Yu.G. Stroganov, Teor. Mat. Fiz. {\bf52} (1982) 105-113
[English trans.: Theor. Math. Phys. {\bf52} (1982) 685-691].
% 3
\bibitem{Hiet1}
J. Hietarinta, private communication
% 4
\bibitem{Hiet2}
J. Hietarinta, {\it Labelling schemes for tetrahedron
equations and dualities between them}, Preprint LPTHE PAR 94/07.
% 5
\bibitem{B}
R.J. Baxter, Commun. Math. Phys. {\bf88} (1983) 185-205.
% 6
\bibitem{Z}
A.B. Zamolodchikov, Zh. Eksp. Teor. Fiz. {\bf79} (1980) 641-664
[English trans.: JETP {\bf 52} (1980) 325-336].
A.B. Zamolodchikov, Commun. Math. Phys. {\bf79} (1981) 489-505.
% 7
\bibitem{BB}
V.V. Bazhanov, R.J. Baxter, J. Stat. Phys. {\bf 69} (1992) 453-485.
% 8
\bibitem{KMS}
R.M. Kashaev, V.V. Mangazeev, Yu.G. Stroganov, Int. J. Mod. Phys. {\bf A8}
(1993) 587-601.
% 9
\bibitem{JM}
M.T. Jaekel, J.M. Maillard, J. Phys. {\bf A15} (1982) 1309.
% 10
\bibitem{StSq}
R.M. Kashaev, V.V. Mangazeev, Yu.G. Stroganov, Int. J. Mod. Phys. {\bf A8}
(1993) 1399-1409.
% 11
\bibitem{NewTw}
V.V. Mangazeev, S.M. Sergeev, Yu.G. Stroganov,
Preprint IHEP 93 -- 126, to appear in Int. J. Mod. Phys.

\end{thebibliography}
\end{document}